\documentclass[aoas,MSNbibl,nameyear,seceqn,dvips]{arximspdf}
\RequirePackage{algpseudocode}
\usepackage{algorithm}
\usepackage{graphicx}

\doi{10.1214/13-AOAS662} %kopijuoti is PTS
\volume{7}
\issue{4}
\pubyear{2013}
\firstpage{2157}
\lastpage{2179}

\makeatletter
\newcommand{\R}{\mathbb{R}}
\newcommand{\D}{\mathcal{D}}
\newcommand{\argmin}{\mathop{\arg\min}}
\newcommand{\calN}{\mathcal{N}}
\newcommand{\X}{\mathbf{X}}
\newcommand{\T}{\mathbf{T}}
\newcommand{\PP}{\mathrm{P}}
\makeatother

\begin{document}
\begin{frontmatter}

\title{Penalized estimation in high-dimensional hidden Markov models with
state-specific graphical~models}
\runtitle{Penalized estimation in high-dimensional HMM{s}}

\begin{aug}
\author[a]{\fnms{Nicolas} \snm{St\"adler}\corref{}\ead[label=e1]{n.stadler@nki.nl}}
\and
\author[a]{\fnms{Sach} \snm{Mukherjee}\ead[label=e2]{s.mukherjee@nki.nl}}
\address[a]{Department of Biochemistry\\
Netherlands Cancer Institute\\
1066CX Amsterdam\\
The Netherlands\\
\printead{e1}\\
\phantom{E-mail:\ }\printead*{e2}}
\affiliation{Netherlands Cancer Institute}
\runauthor{N. St\"adler and S. Mukherjee}
\end{aug}

% HISTORY:
\received{\smonth{9} \syear{2012}}
\revised{\smonth{5} \syear{2013}}

% ABSTRACT
%
\begin{abstract}
We consider penalized estimation in
hidden Markov models\break  (HMMs) with multivariate Normal observations.
In the moderate-to-large dimensional setting, estimation for HMMs
remains challenging in practice, due to several
concerns arising from the hidden nature of the states.
We address these concerns by $\ell_1$-penalization of state-specific
inverse covariance matrices.
Penalized estimation leads to sparse
inverse covariance matrices which can be interpreted as state-specific
conditional independence graphs.
Penalization
is nontrivial in this latent variable setting;
we propose a penalty that automatically adapts to the number of states
$K$ and
the state-specific sample sizes and can cope with scaling issues arising
from the unknown states.
The
methodology is adaptive and very general, applying in particular to
both low- and high-dimensional settings without
requiring hand tuning.
Furthermore, our approach facilitates exploration of the number of
states $K$
by coupling estimation for successive candidate values $K$.
%Furthermore, it remains difficult to determine an appropriate number
%of states in practical problems.
%: There are no restrictions on sample size,
%neither on number of covariates and neither on number of hidden
%states. In particular, it also covers the %There are no restrictions
%on sample size in particular, it also covers the
%high-dimensional scenario, with small sample-size compared to a large
%parameter
%space. Our methodology involves optimization of the $\ell_1$-penalized
Empirical results on simulated examples demonstrate
the effectiveness of the proposed approach.
%that the proposed methodology can substantially outperform both
%classical, unpenalized and naively penalized estimators, even when the
%sample size is seemingly large.
In a challenging real data example from genome biology,
we demonstrate the ability of
our approach to yield gains in predictive power and to deliver richer
estimates than existing methods.
%and sheds new light on open questions in the scientific application
%domain.

%Our framework is able to deal with several difficulties arising from
%the unknown nature of the hidden states.
%Chromatin modeling \citep{filion2010,kharchenko2011} serves as our
%prime motivation and illustration. Chromatin is the
%ensemble of DNA and proteins which make up the nucleus of a eukaryotic
%cell. The task in chromatin modeling is to segment the genome into
%different regions (or states in our glossary) based on protein binding.
%%It is
%%thought that protein binding is very specific to different regions of
%%the genome and thats how chromatin regulates overall gene
%%expression.
%We apply our method to genome-wide binding data of 53 proteins. We are
%able to infer the number of states, we provide state-specific binding
%levels and importantly, our method also gives state-specific networks
%indicating how the different proteins interplay within different
%states.

%HMM\&GLasso for joint segmentation and network inference.

%Tuning Parameter: choice of K (nr of states), $\lambda$ in high-dim
%setup \& initialization
\end{abstract}

% KEYWORDS

\end{frontmatter}

%s1 #&#
\section{Introduction}\label{sec1}
In this paper we consider estimation in high-dimensional hidden Markov
models. % \citep{rabiner1989}.
We consider multivariate observations $X_t \in\R^p$ with discrete
index $t \in\mathcal{T} = \{ 1, \ldots, n \}$ and hidden states
$S_t \in\{1 ,\ldots, K \}$ (the models we consider are high-dimensional
in the sense of relatively large $p$).
Conditional on state, emission distributions are multivariate Normal
(MVN), with $X_t \mid S_t = k \sim\mathcal{N}(\mu_k,\Sigma_k)$ [where
$\mathcal{N}(\mu,\Sigma)$ denotes the MVN density with mean $\mu$ and
covariance matrix $\Sigma$].
%HMMs are widely used in diverse applications, including, among others,
%speech recognition \citep{rabiner1989}, medical signal processing and
%genomics.
Estimation in the small $p$ case of univariate or low-dimensional
observations is a well-studied problem.
In contrast, estimation in the larger $p$ setting remains challenging
due to several factors:
\begin{longlist}[(iii)]
\item[(i)]\textit{High-dimensionality.} Inference in HMMs with moderate or
large number of features is, in a sense, always a high-dimensional
problem since the ratio $\min_k n_k/p$
may be small, as it depends on the \emph{unknown} number of states
and the \emph{unknown} size of the states ($n_k$ denotes the number of
samples in state $k$). Therefore, large samples for each state cannot
be relied upon at the outset, even when the overall sample size $n =
\sum_k n_k$ is large.
\item[(ii)]\textit{Covariance structure.} Estimation is especially
challenging in settings where covariances $\Sigma_k$ cannot be assumed
to have a simple structure (e.g., diagonal) or where state-specific
covariance structure is itself of scientific interest. Then, due to
Simpson's paradox, state-specific covariances must be jointly estimated
along with state assignments.
\item[(iii)]\textit{Regularization.} The size and scale of individual states
may vary and are usually unknown at the outset. Regularization schemes
need to self-adapt appropriately.
\item[(iv)]\textit{Number of hidden states.} The model selection problem of
determining or exploring the number of states $K$ is coupled to the
estimation problem for known~$K$. In the multivariate setting,
estimation for known $K$ is itself challenging. Then, the
straightforward strategy of fitting models for various values $K$ and
comparing by model selection criteria may become difficult or intractable,
especially when practically important issues like initialization and
setting of tuning parameters are taken into consideration.
\end{longlist}

This work is motivated by
applied questions in genome biology; we present below a real data
example from that field.
HMMs are very widely used in genomics.
Measurements at genome locations $t$ constitute the vector $X_t$, while
states $S_t$ are typically identified with
biological states of the genome (e.g., whether the location~$t$ is
within a gene-coding region).
%The sample size $n$ is the size of the genome (or part thereof) and is
%therefore typically large.
Early, pioneering applications of HMMs to genomic data [see, e.g.,
\citet
{krogh,durbin1998}] considered univariate or low-dimensional
observations $X_t$ (such as the gene sequence itself).
However, in recent years technological advances have begun to permit
higher dimensional studies.
For example, using technologies such as DamID [\citet
{vansteensel2000identification}] or ChIP-seq [\citet{park2009chip}], it
is now possible to measure the binding of proteins to the DNA across
the entire genome for dozens or hundreds of proteins and the
dimensionality (i.e., number of proteins) of such approaches continues
to increase; see, for example, \citet{encode}.
Gene expression depends not only on sequence (the genome) but also on
a diverse set of regulatory mechanisms including the binding of
protein transcription factors to the DNA. Protein-DNA binding can be
influential in regulating transcription, for example, cells belonging
to different tissue types in the same organism (with the same genome)
may have quite different protein-DNA binding patterns, expression
profiles and biological functions.
The importance of protein-DNA binding in understanding such epigenetic
variation has led to much interest in studying the genome in terms of
binding patterns and in identifying regions of the genome with shared
regulatory influences.
At present such analyses are performed using HMMs
where the states $S_t$ are identified with biological states and
observations with multivariate protein-DNA binding data [\citet
{filion2010,ernst2010}].
However, absent reliable methodology for fitting high-dimensional HMMs,
it is common practice in the field to instead consider reduced
dimension versions of the data [by selecting key ``marker'' variables or
carrying out dimensionality reduction as a preprocessing step; see,
e.g., \citet{filion2010}] or by discretizing the data and treating
observations as independent Bernoulli [\citet{ernst2010}]. We show below
in a real data example from genome biology that
our penalized approach applied to all available variables (proteins)
from a recent experiment yields large gains in predictive accuracy (on
held-out test data) relative to a reduced-dimension approach, as well
as relative to classical estimation applied to the full set of variables.
Beyond genomics, potential application areas for high-dimensional HMMs
are diverse and include biomedical signal processing (e.g., analysis of
multi-channel EEG data), engineering applications (including image and
video processing) and finance.

% what do we solve? how is it better
% what do we actually do: use penalisation,
%
%We propose new methodology for estimation in HMMs with multivariate
%observations.
We propose a
penalized log-likelihood procedure involving $\ell_1$-norms
of the state-specific inverse covariance matrices $\Sigma_k^{-1}$, with
optimization carried out within an expectation-maximization (EM) framework.
Our approach has several attractive features:
\begin{itemize}
\item Penalized estimation leads to sparse
inverse covariance matrices which can be interpreted as state-specific
conditional independence graphs or networks [\citet
{yuan05model,friedman2007sic}].
\item The specific penalty we propose automatically adapts to the
number of states
and state-specific sample size and enjoys scale invariance that takes
care of state-specific scaling.
\item The number of states $K$ can be selected automatically, or
estimates for various values $K$ explored,
using a computationally efficient approach that couples
estimation for successive candidate values for $K$.
\item The approach requires essentially no hand tuning;
only a maximum number of states $K_\mathrm{max}$ must be set by the user.
Otherwise, tuning parameters (including, if desired, $K$ itself) are
set automatically.
% and is robust with respect to initialization.
\end{itemize}

Our approach is very general: as we demonstrate below, it works well in
diverse regimes, including both low- and high-dimensional examples,
with no hand-tuning required. In a real data example from genomics
the methodology leads to large gains in predictive power relative to
existing approaches.

% WHY IS PENALIZATION SUBTLE?
% penalization can be done in an obvious way. but the details mater,
%the form of the penalty, etc.
Penalized estimators can be incorporated into EM-type algorithms and a
number of recent authors have done so, notably in the context of
mixture models [\citet{khalili,fmrlasso2009,pan,hill2013}]. However, the
unknown nature of the states (or mixture components) poses special
challenges for penalization that have not been adequately addressed so far.
In particular, appropriate penalization must account for the
number of hidden states and their respective sample sizes, but these are
themselves unknown at the outset. Furthermore, scaling also poses a
subtle problem: in the
classical Lasso [\citet{tibshirani96regression}] or Graphical Lasso
[\citet{friedman2007sic}] standardization is an important preprocessing
step to ensure appropriate scaling.
However, in HMMs and mixtures different states or components may differ
with respect to scale, but since state assignments are {a
priori} unknown, standardization cannot be carried out as a
preprocessing step. The penalty we propose automatically adapts with
state-sizes and takes care of scaling issues.
Inspired by the seminal paper of \citet{donoho94} and related work in
the Lasso context [\citet{zhang2010,sun2011,barron2008}],
our penalty allows for \emph{universal regularization} by use of a
tuning parameter $\lambda_\mathrm{uni}$, that depends only on $n$ and $p$.
Using universal regularization by $\lambda_\mathrm{uni}$ within our EM
algorithm allows automatic adaptation to number of states $K$ and
state-specific sample sizes.
As a consequence of these features, our procedure for penalized
estimation for a given number of states $K$ is entirely free of
user-set parameters.

%We use l1. but why is it subtle and what do we do?

% WHAT IS OU MODEL SELETION?
%Brute-force determination of $K$, by separately fitting models for
%various $K$ and subsequent comparison by model selection criteria,
%does not scale well, since estimation for a given $K$ is already
%computationally demanding in practice. However, parameter estimates
%for successive values $K, K+1$ are related, and it is therefore
%natural to
Parameter estimates for successive values $K, K+1$ are related, and it
is therefore natural to
exploit this fact in exploring the number of states;
we do so using an iterative algorithm. In principle, an iterative
approach could proceed in a ``top down'' manner from few states to many,
or ``bottom up'' from many states to few.
However, we cannot in general gain information about two underlying
states from
estimates obtained from a single, merged state (Simpson's paradox);
this means the ``top down'' approach cannot be reliably used in the
multivariate setting.
We therefore proceed in a ``bottom up'' manner, starting with a large
number of states $K_\mathrm{max}$
and iteratively reducing the number of states through the entire
considered range.
Model order reduction is guided using the Kullback--Leibler divergence
between state densities; this naturally takes account of both mean and
covariance information.
This exploration is efficient because (i) current estimates are used to
provide initialization for the subsequent iteration
and (ii) we initialize the EM algorithm only once, at the first
iteration corresponding to $K=K_\mathrm{max}$. As we demonstrate below,
this procedure in fact outperforms the ``brute-force''
approach of entirely separately fitting models for various $K$'s. In
this way, our approach allows
tractable exploration of estimates for a range of values $K$ and, if
desired, automatic selection of $K$.
Our approach is inspired by the work of
\citet{Figueiredo2000} who used a similar
strategy in the context of low-dimensional mixtures.
\section{Inference in hidden Markov models with state-specific
graphical models}\label{sec:method}

We consider a hidden Markov model (HMM) with multivariate Normal (MVN)
emissions. We denote by
$S_t \in\{1,\ldots,K\}$ the (hidden) state process, that is, a discrete
Markov chain with transition matrix
${\Pi}_{kk'}=\PP(S_{t+1}=k'|S_t=k)$; in order to simplify the
notation, we omit the initial probabilities $p_{k}=\PP(S_1=k)$ in the
further description of our methodology.
We denote by $X_t\in\R^{p}$ the observed process with emission
distribution $X_t \mid S_t=k\sim\calN(\mu_k,\Sigma_k)$.

The case of sparse inverse covariance matrices $\Omega_k = \Sigma
_k^{-1}$ will be of particular interest.
For each state we have a Gaussian graphical model with undirected graph
$G_k$ defined by locations of zero entries in the inverse covariance
matrix, that is, $(l,l') \notin G_k \iff(\Omega_k)_{ll'} = 0$.
%
% between $X^l$ and $X^{l'}$ in state k} if and only if $(
We denote model parameters by
$\Theta_K=(\theta_1,\ldots,\theta_K,\Pi), \theta_k=(\mu_k,\Omega_k)$.
The goal, for given $K$, is to infer $\Theta_K$ from the observed
$n\times p$ data
matrix $\X$, and further to solve the related problem of exploring (or
determining) $K$ itself.

Conceptually, it makes sense to think of inference in a HMM (or mixture model)
%with state-specific Gaussian graphical models
as a combination of two (coupled) tasks.
The first task consists of estimating the model parameter $\Theta_K$,
%%=(\theta_1,\ldots,\theta_K,\Pi)$, $\theta_k=(\mu_k,\Omega_k)$
given the number of states $K$ and a regularization parameter~$\lambda$.
% (and also given initialization required as an input for the
%algorithm).
For this task, we propose to minimize the negative penalized log-likelihood
%
%e2.1 #&#
\begin{equation}
\label{eq:mpl} \hat{\Theta}_{K,\lambda}=\argmin_{\Theta_{K,\lambda}}-\ell (\Theta
_{K,\lambda};\X)+\lambda\operatorname{pen}(\Theta_{K,\lambda}),
\end{equation}
where $\ell(\Theta_{K,\lambda};\X)$ denotes the observed
log-likelihood and $\operatorname{pen}(\Theta_{K,\lambda})$ is a penalty
function involving the $\ell_1$-norms of the inverse covariance
matrices [\citet{yuan05model,friedman2007sic,meinshausen04consistent}]
that we describe in detail below.
%In this first task we solve (\ref{eq:mpl}) for fixed $K$ and $
%when $p$ is large in comparison to state-size.
The $\ell_1$-norm is especially appealing when the goal is
network inference, as it induces sparsity in
$\Omega_k$'s and therefore in the corresponding undirected graphs $G_k$.
We solve this problem by an EM-type algorithm,
using a specific penalty that we describe below;
we call this approach \emph{HMMGLasso} (see Section~\ref{sec:hmmglasso} for details).
The adaptive regularization strategy we propose in HMMGLasso permits
estimation of HMMs with state-specific covariance structure in both
low- and high-dimensional settings, while taking care of state size and
scaling; this addresses points (i)--(iii) raised in the \hyperref[sec1]{Introduction}.
%We note that as with every EM
%algorithm, HMMGLasso only converges to a local optimum and depends on
%initialization.

%the choice of initial
%parameters. Therefore, the solution to (\ref{eq:mpl}) depends not only
%on $K$ and $\lambda$ but also on initialization.

The second task
involves determining
an appropriate
number of states
$K^*$ and suitable penalization parameter $\lambda^*$.
This is a model selection problem, and can in principle be solved by
minimizing a model selection criterion $\mathcal{C}(K,\lambda)$ (we
consider specific criteria below), that is,
%
%e2.2 #&#
\begin{equation}
\label{eq:mincrit} \bigl(K^*,\lambda^*\bigr)=\argmin_{K,\lambda}
\mathcal{C}(K,\lambda).
\end{equation}
%
%Bic is a very popular selection criterion. The MMDL (mixture minimum
%description length) an adaptation of Bic for mixture models is another
%one (see Section).
%Below we consider the Bayesian information criterion (BIC) and mixture
%minimum description length (MMDL) as candidate criteria $\mathcal{C}$.

%It is important to recognize that the problems (\ref{eq:mpl}) and (
%needed at each grid point to guard against local optima.

As described in detail below, we propose an iterative approach called
\emph{Greedy Backward Pruning} that exploits the relationship between
estimates $\hat{\Theta}_K$ for successive $K$'s to allow efficient
model exploration and, if desired, determination of~$K$. This addresses
point (iv) raised in the \hyperref[sec1]{Introduction}.
Using Greedy Backward Pruning, initialization is carried out once at a
(too) large
number of states $K_\mathrm{max}$; as we show below, this strategy gives
highly competitive estimates despite needing only a single initialization.

\subsection{HMMGLasso in detail: Baum--Welch algorithm and $\ell_1$
regularization}\label{sec:hmmglasso}
%First describe parameter estimation in HMM with MVN-emissions (to
%introduce necessary notation). Then show how to adapt/regularize for
%network inference.
Maximum likelihood estimation for HMM is usually performed using the EM
algorithm (or the Baum--Welch algorithm in the HMM context).
%Let $\ell_{c}(\Theta;\X,\mathbf{S})$%=\ell_{c}(\mu_{k=1,\ldots,K},
%log-likelihood which can be written as
Denote the complete log-likelihood with
\[
\label{eq:loglik.c} \ell_{c}(\Theta;\X,\mathbf{S})=\sum
_k\ell\bigl(\mu_k,\Omega_k;\T
^k_1,\T ^k_2\bigr)+\ell(\Pi;
\T_3),
\]
where
$\mathbf{S} = (S_1, \ldots, S_n)$ are state assignments, $\mathbf{X} =
(X_1, \ldots, X_n)^{\mathrm{T}}$ is the $n \times p$ data matrix,
$\ell(\mu_k,\Omega_k;\T^k_1,\T^k_2)$ is the log-likelihood of the MVN
distribution with mean
$\mu_k$ and inverse covariance $\Omega_k$ and $\ell(\Pi;\T_3)$ is the
log-likelihood
of the Markov chain with transition matrix $\Pi$. $\T^k_1 =\sum_{t}\mathbf{1}_{(S_t=k)}X_t, \T^k_2 =\sum_{t}\mathbf{1}_{(S_t=k)}X_tX_t^{T}$
and $(\T_3)_{kk'}=\sum_{t}^{}\mathbf{1}_{(S_{t}=k,S_{t+1}=k')}$
are the corresponding sufficient statistics.

% CAN WE SHORTEN THE BELOW?
Following initialization, EM produces a sequence of estimates $\{\Theta
^{(i)};i=1,2,3,\ldots\}$ by alternating between E- and M-Steps.
% \begin{itemize}
% \item E-Step: Compute conditional expectation of complete
% log-likelihood, given observed data $\X$ and current estimate
% $\Theta^{(i)}$. This involves:
% \begin{eqnarray*}
% \mathrm{u}_k^{(i)}(t)&=&\PP_{\Theta^{(i)}}(S_t=k|\X)\\
% \mathrm{v}_{kk'}^{(i)}(t)&=&\PP_{\Theta^{(i)}}(S_t=k,S_{t+1}=k'|\X)
% \end{eqnarray*}
% These quantities can be efficiently computed using the
%Forward-Backward equations. The expected complete log-likelihood equals
% \begin{eqnarray}
% \label{eq:e.loglik.c}
% \E_{\Theta^{(i)}}[\ell_c(\Theta;\X,\mathbf{S})]&=&\sum_{k}\ell(\mu_k,
% \end{eqnarray}
% where
% $\T^{u^{(i)}_k}_1=\sum_{t}u^{(i)}(t)X(t)$, $\T^{u^{(i)}_k}_2=\sum
%_{t}u^{(i)}(t)X(t)X(t)^{T}$
% and $\T^{v^{(i)}}_3=\mathrm{v}^{(i)}$.
% \item M-Step: Maximize the expected log-likelihood w.r.t. $\Theta$ to
% obtain:
% \begin{eqnarray}
% \label{eq:mstep}
% \Theta^{(i+1)}=\argmax_\Theta\E_{\Theta^{(i)}}[\ell_c(\Theta;\X,
% \end{eqnarray}
% In light of (\ref{eq:e.loglik.c}) the M-Step
% can be performed separately w.r.t. $\Pi$ and w.r.t. each
%state-specific parameters
% $(\mu_k,\Sigma_k)$.
% \end{itemize}
To facilitate network inference, we seek to induce sparsity in the
$\Omega_{k}$'s. We do this by $\ell_1$-regularization. In particular,
we replace maximization with respect to $(\mu_k,\Omega_k)$ in the
M-Step of the Baum--Welch algorithm by
%
%e2.3 #&#
\begin{equation}
\label{eq:2}\qquad \bigl(\mu_{k}^{(i+1)},\Omega_{k}^{(i+1)}
\bigr)=\argmin_{\mu_k,\Omega_k} -\ell\bigl(\mu_k,\Omega_k;
\T^{u^{(i)}_k}_1,\T^{u^{(i)}_k}_2\bigr)+\lambda
\sqrt{\pi _k^{(i)}}\operatorname{Pen}(
\Omega_k).
\end{equation}
Here,
\[
\T^{u^{(i)}_k}_1 =\sum_{t}u_k^{(i)}(t)X_t,\qquad
\T^{u^{(i)}_k}_2 =\sum_{t}u_k^{(i)}(t)X_tX_t^{T}
\]
denote the expected sufficient statistics given $\X$ and current estimate
$\Theta^{(i)}$ with state-responsibilities
$\mathrm{u}_k^{(i)}(t)=\PP_{\Theta^{(i)}}(S_t=k|\X)$ obtained from the
E-Step.

By $\pi_k^{(i)}=n_k^{(i)}/n$ ($n_k^{(i)} = \sum_t u_k^{(i)}(t)$) we
denote the (scaled) effective sample size
of state $k$. The penalty term depends on a regularization parameter
$\lambda$, on the effective sample size $\pi_k^{(i)}$ and on a
function $\operatorname{Pen}(\cdot)$ involving $\ell_1$-norm of
$\Omega_k$. The reason why we incorporate the square root of the
effective sample size is that it is known from the Lasso literature that
the $\ell_1$-penalty term asymptotically has to grow with the square root
of the sample size in order to achieve optimality [\citet
{lassobook2011}]. We consider three slightly different functions
$\operatorname{Pen}(\cdot)$ defined as follows:
\begin{itemize}
\item$\operatorname{Pen}_\mathrm{invcov}(\Omega)=\|\Omega^-\|_1$, the classical
penalty known from the Graphical Lasso. It imposes $\ell_1$-constraints
on the nondiagonal entries of the concentration matrix~$\Omega$.
\item$\operatorname{Pen}_\mathrm{parcor}(\Omega)=\|\Psi^-\|_1$, where $\Psi
$ is
the partial correlation matrix which can be written as $(\Psi
)_{ll'}=-\Omega_{ll'}/\sqrt{\Omega_{ll}\Omega_{l'l'}}$.
\item$\operatorname{Pen}_\mathrm{invcor}(\Omega)=\|\Phi^-\|_1$, where $\Phi
$ is
the inverse of the correlation matrix given by $\Phi=\mathrm{C}^{-1},
\mathrm{C}_{ll'}=\Sigma_{ll'}/\sqrt{\Sigma_{ll}\Sigma_{l'l'}}$.
\end{itemize}
Note that all three functions penalize the $\ell_1$-norm of the
concentration matrix and therefore lead to sparse $\Omega$'s. The
advantage of $\operatorname{Pen}_\mathrm{parcor}(\cdot)$ and $\operatorname
{Pen}_\mathrm{invcor}(\cdot)$ is that
%they do not depend on scaling of the input data and can in principle
%address the concern
they are scale-invariant and therefore remove concerns that arise from
state-specific scaling.
As we noted above, state-specific scaling cannot be removed by
preprocessing in the HMM setting since
state assignments are themselves unknown at the outset.
% We believe that scale-invariance is a critical point in HMMs as
%hidden states and parameters are inferred simultaneously and
%state-specific scaling as a preprocessing step is not possible.

Optimization of (\ref{eq:2}) is nonstandard. Noting that
\begin{eqnarray*}
\ell\bigl(\mu_k,\Omega_k;\T^{u^{(i)}_k}_1,
\T^{u^{(i)}_k}_2\bigr)&=&\frac
{n_k^{(i)}}{2}\log |\Omega_k|-\frac{1}{2}\operatorname{tr}\bigl(\Omega_k \T
^{u^{(i)}_k}_2\bigr)+\mu_k^{T}
\Omega_k \T^{u^{(i)}_k}_1\\
&&{}-\frac{1}{2}n_k^{(i)}
\mu_k^{T} \Omega_k \mu_k,
\end{eqnarray*}
it is easy to verify that (\ref{eq:2}) reduces to $\mu_k^{(i+1)}=\T
^{u^{(i)}_k}_1/n_k^{(i)}$,
%
%e2.4 #&#
\begin{equation}
\label{eq:glasso}\quad  \Omega_{k}^{(i+1)}=\argmin_{\Omega_k}-
\log|\Omega_k|+\operatorname {tr}\bigl(\Omega_k
\mathbf{C}^{u^{(i)}_k}\bigr)+2\frac{\lambda
}{n_k^{(i)}}\sqrt{\pi
_k^{(i)}} \operatorname{Pen}(\Omega_k),
\end{equation}
where
$\mathbf{C}^{\mathrm{u}^{(i)}_k}=\frac{1}{n_k^{(i)}}\T
^{u^{(i)}_k}_2-\mu
_k^{(i+1)}(\mu_k^{(i+1)}) ^{T}$. For the penalty function $\operatorname
{Pen}_\mathrm{invcov}(\cdot)$ optimization problem (\ref{eq:glasso}) can
be solved by the Graphical
Lasso algorithm presented in \citet{friedman2007sic}. In the
supplementary material [\citet{supp}] we compare these three different
penalties and discuss how we perform optimization.

\begin{algorithm}
\caption{HMMGLasso}\label{alg1} % enter the algorithm environment
\begin{algorithmic}[1] % give the algorithm a caption
% enter the algorithmic environment
\State\textbf{Input} $K, \lambda, \Upsilon^{(0)}\,{=}\,\{\!(\mathrm
{u}^{(0)}_k(t)\!)_{k=1,\ldots,K,t \in
\mathcal{T}},\Pi^{(0)},\pi^{(0)}\!\}$ and set $i\,{=}\,0,\mathrm{err}^{(0)}\,{=}\,0$.

\While{$\{\mathrm{err}^{(i)}>\varepsilon\}\vee\{\pi^{(i)}_{k}>\pi
_\mathrm{min} \mbox{ for all } k=1,\ldots, K\}$}%\vspace{0.1cm}

\State\textbf{M-Step} Obtain estimates

$(\mu_k^{(i+1)},\Omega_k^{(i+1)})=\argmin_{\mu_k,\Omega_k}-\ell
(\mu
_k,\Omega_k;\T^\mathbf{u_k^{(i)}}_1,\T^\mathbf
{u_k^{(i)}}_2)+\lambda
\sqrt{\pi^{(i)}_k}\times\break \hspace*{20pt}\operatorname{Pen}(\Omega_k)$

$\Pi^{(i+1)}_{kk'}=\mathbf{v}^{(i)}_{kk'}/\pi^{(i)}_k$
($\Pi^{(1)}_{kk'}=\Pi^{(0)}_{kk'}$ in 1st iteration) %\vspace{0.1cm}

\State\textbf{E-Step} Use Forward-Backward equations to update

$\mathrm{u}^{(i+1)}_k(t)=\PP_{\Theta^{(i+1)}}(S_t=k|\X)$%\vspace{0.2cm}

$\mathrm{v}^{(i+1)}_{kk'}(t)=\PP_{\Theta
^{(i+1)}}(S_t=k,S_{t+1}=k'|\X
)$%\vspace{0.2cm}

$\pi^{(i+1)}_k=\sum_t\mathrm{u}^{(i+1)}_k(t)/n$
\State\textbf{Set} $\mathrm{err}^{(i+1)}=\max_{k,l,l'} \{\frac
{|\Sigma_{k,ll'}^{(i+1)}-\Sigma_{k,ll'}^{(i)}|}{1+|\Sigma
_{k,ll'}^{(i+1)}|} \}$ and
$i\leftarrow i+1$
\EndWhile%\vspace{0.1cm}
\State\textbf{Output} $\hat\Xi^{(K,\lambda)}=\{\hat\Theta
_{K,\lambda
}, (\mathrm{\hat u}_k(t) )_{k=1,\ldots,K,t \in
\mathcal{T}},\hat{\pi}\}$
\end{algorithmic}
\end{algorithm}

Algorithm~\ref{alg1} summarizes HMMGLasso. As stated above, the EM algorithm
depends on initial specification of parameters, that is, $\theta
_k^{(0)},\Pi^{(0)}$ ($k=1, \ldots,K$). For convenience
(see later in text) we directly specify $\mathrm{u}_k^{(0)}(t)$
(instead of $\theta_k^{(0)}$) and start with an M-Step
followed by an E-Step. We stop the algorithm if the relative change in
the $\Sigma_k$'s falls below a threshold $\varepsilon$ or if for at least
one state the scaled effective sample size $\pi_k$ is smaller than
$\pi
_\mathrm{min}$.

\subsection{Universal regularization}\label{sec:launi}

In this section we discuss the choice of the regularization parameter
$\lambda$ in HMMGLasso. We will argue that $\lambda_{\mathrm
{uni}}=\sqrt {2n\log p}/2$ is a
reasonable regularization parameter for HMMGLasso. We do this by considering
connections with the Lasso [\citet{tibshirani96regression}] and
the Graphical Lasso [or GLasso; \citet{friedman2007sic}]. In the
classical Lasso or
GLasso setup the regularization parameter is usually chosen empirically
to minimize the prediction error (e.g., by performing
cross-validation). However, in the HMM (or more generally latent
variable) setting, with unknown number of states $K$,
such a brute force strategy is computationally burdensome, motivating
the need for universal regularization.

%% and by using
%% theory on the Lasso which indicates the amount of penalization
%% necessary to obtain optimality results (see for example...).
First, consider a classical regression setup with
$y=\X\beta+\varepsilon$, where
$\varepsilon\sim\calN(0,\sigma^2\mathrm{I})$.
Here, $\X$ is a $N\times p$ predictor matrix, $y$ a $N\times1$
response vector, $\beta$ denotes the $p\times1$
regression parameter and $\sigma^{2}$ is the error variance. Then, the
Lasso estimator minimizes %is defined by
$\|y-\X\beta\|^{2}/2N+s\|\beta\|_1$. Assuming an orthonormal predictor
matrix, \citet{donoho94} showed that the risk of the
Lasso estimator comes close
to the oracle risk if we use $s_\mathrm{uni}=\sigma\sqrt{2 \log p/N}$ as
a regularization parameter. Universal regularization and the penalty
$\sigma\sqrt{2
\log p/N}$ are discussed also in the nonorthonormal case in \citet
{zhang2010} or \citet{sun2011} [see
also \citet{barron2008}; they propose a universal penalty parameter
based on the minimum description length principle]. It is important to
note that $s_{\mathrm{uni}}$ decreases with $1/\sqrt{N}$. This is the
reason why we include the square-root of the effective sample size into
the state-specific penalty terms in the HMMGLasso~(see Section~\ref{sec:hmmglasso}).

Next, consider the Graphical Lasso,
\[
\hat{\Omega}=\argmin_{\Omega} -\log|\Omega|+\operatorname {tr}(\mathbf{S}
\Omega) + \rho\bigl\|\Omega^-\bigr\|_1,
\]
where $\mathbf{S}$ is the sample covariance matrix of
$X=(X^{(1)},\ldots,X^{(p)})\sim\calN(0,\Sigma)$ with $\Omega
=\Sigma^{-1}$.
\citet{friedman2007sic} showed that the last row/column of $\hat\Omega$
can be obtained by solving
%
%e2.5 #&#
\begin{equation}
\label{eq:lassoglasso} \hat{\beta}=\argmin_{\beta} 0.5\beta{
\Sigma_{11}}\beta -\beta\mathrm{s}_{12}+\rho\|\beta
\|_1,
\end{equation}
where $\beta$ and $\Omega$ are linked through
$\sigma_{12}=\Sigma_{11}\beta$ ($\Sigma_{11}$ is the covariance
matrix with the last row and column deleted; $\sigma_{12}$ and
$\mathrm{s}_{12}$ denote the last row of the covariance and sample
covariance matrix). Note that (\ref{eq:lassoglasso}) can be interpreted
as the Lasso
estimator corresponding to regression of variable $X^{(p)}$ against
$X^{(1)},\ldots,X^{(p-1)}$. As $1/\Omega_{pp}$ is the error variance in
regressing $X^{(p)}$ against $X^{(1)},\ldots,X^{(p-1)}$, we can identify
${\Omega_{pp}^{-1/2}} \sqrt{2\log p/N}$ as a good choice for
$\rho$ in (\ref{eq:lassoglasso}). If $\Omega$ is
standardized to have unit diagonal entries, then we can write $\rho
_\mathrm{uni}=\sqrt{2\log p/N}$.

Now consider equation (\ref{eq:glasso}) of the HMMGLasso with
$\operatorname{Pen}_\mathrm{invcov}(\cdot)$ and assume all
$\Omega_k$'s are standardized to have unit diagonal. Equating
$2\frac{\lambda}{n_k}\sqrt{\pi_k^{(i)}}$ with $\rho_\mathrm{uni}=\sqrt {2\log p/n_k}$
(the universal shrinkage level in the Graphical Lasso
with sample size $N=n_k$) and solving for $\lambda$, we obtain
\[
\lambda_{\mathrm{uni}}=\sqrt{2n\log p}/2.
\]

For the penalty function $\operatorname{Pen}_\mathrm{invcov}(\cdot)$ the foregoing indicates that
$\lambda_\mathrm{uni}=\sqrt{2n\log p}/2$ only holds if the $\Omega_k$'s are
standardized and therefore equal the corresponding partial correlation
matrix. In general, since state assignments are themselves unknown,
this standardization cannot be done as a preprocessing step.
However, if we use $\operatorname{Pen}_\mathrm{parcor}(\cdot)$ instead,
$\lambda_\mathrm{uni}=\sqrt{2n\log p}/2$ applies regardless of scaling. Penalizing
the partial correlation can be seen as a generalization of the
``scaled'' Lasso proposed by \citet{fmrlasso2009}. There, the negative
log-likelihood is penalized by $s\frac{\|\beta\|_{1}}{\sigma}$ and
optimization is performed over $\beta$ and $\sigma$ simultaneously. A
reasonable choice for $s$ is $\sqrt{2 \log p/N}$, which does not
depend anymore on the unknown noise level [see \citet{sun2011} and
also the discussion in \citet{fmrlasso2009}].

Thus, $\lambda_\mathrm{uni}$ is the penalty level we use for estimation
in HMMGLasso.
%It is important to note that $\lambda_\mathrm{uni}$ does not depend on
%the number of states $K$.
It is ``universal'' in the sense
that it only depends on the dimensionality of the input data $n$ and
$p$. Furthermore, when $\lambda_\mathrm{uni}$ is used with the penalty
$\operatorname{Pen}_\mathrm{parcor}(\cdot)$ the penalization self-adapts to the hidden states by
incorporating the
square-root of the effective sample size and by taking care of
scaling.

%s2.3 #&#
\subsection{Model order exploration using Greedy Backward Pruning}
\label{sec:agglo}
% Criteria: BIC, MMDL}\label{sec:bicmmdl}
%In this Section we describe in detail the Greedy Backward Pruning
%algorithm.
Greedy Backward Pruning can in principle be used with a wide range of
model selection criteria; here we consider
the popular Bayesian Information Criterion (BIC) and the Mixture
Minimum Description Length (MMDL). MMDL was introduced by
\citet{Figueiredo1999} and was specifically proposed for the purpose of
determining the
number of components in finite mixtures.
We first describe these criteria and then go on to give a detailed
description of the Greedy Backward Pruning algorithm.

%pa2.3.subsubsection.1 #&#
\textit{Model selection criteria.}
A model selection criterion $\mathcal{C}$ has to trade off
goodness of fit and model complexity. BIC and MMDL are defined by
\begin{eqnarray*}
\operatorname{BIC}(\hat\Theta_{K,\lambda})&=&-\ell(\hat{\Theta }_{K,\lambda};\X )+
\frac{1}{2}\log(n)K(K-1)+\frac{1}{2}\log(n)\sum
_k\operatorname {Df}(k,\lambda),
\\
\operatorname{MMDL}(\hat\Theta_{K,\lambda})&=&-\ell(\hat{\Theta }_{K,\lambda};\X
)+\frac{1}{2}\log(n)K(K-1)+\sum_k
\frac{1}{2}\log(n\hat\pi _k)\operatorname {Df}(k,\lambda),
\end{eqnarray*}
where in the context of $\ell_1$ penalized log-likelihood we set the
degrees of freedom as
$\operatorname{Df}(k,\lambda)=p+\sum_{l'\geq l}\mathbf{1}_{(\hat\Omega
_{k,\lambda})_{ll'}\neq
0}$.

MMDL can be motivated by the minimum description length principle
[\citet{Gruenwald2007}]. The negative log-likelihood represents the
optimal code-length of the data given model parameters
$\Theta$. The term $\frac{1}{2}\log(n)K(K-1)$ is the
``optimal'' code-length for the transition matrix $\Pi$ (note that
$\Pi
$ is
estimated from all data). As $n\pi_k$ is the effective sample size
from which
$\theta_k=(\mu_k,\Omega_k)$ is estimated, we get
$\frac{1}{2}\log(n\pi_k)\operatorname{Df}(k,\lambda)$ as an ``optimal''
code-length for describing $\theta_k$.

The main difference between BIC and MMDL is the use of the
\emph{effective sample size} $n\hat{\pi}_k$ in the code-lengths for
parameters which are state-specific. \citet{Figueiredo1999} argued
using ideas from minimum description length literature that MMDL is more
appropriate for mixtures than BIC. They demonstrate on real and
synthetic data that MMDL
outperforms BIC. In Section~\ref{sec:numexp} we
compare performance of Greedy Backward Pruning using BIC and MMDL
as model selection criteria. In our more involved inference task
we come to the same conclusion as \citet{Figueiredo1999}, namely, that
MMDL outperforms BIC.

%Especially in a
%high-dimensional setting (with large $p$ and small $n$) the gains of
%using MMDL are noticeable. It seems that in such scenarios the penalty
%term in BIC and MMDL is much more sensible than in the classical $n$
%large, $p$ small setting.

%pa2.3.subsubsection.2 #&#
\textit{Greedy Backward Pruning in detail}.
Greedy Backward Pruning works by first
estimating parameters using HMMGLasso
with a large
number of states $K_\mathrm{max}$ %(this should be larger than the correct
%number of states)
and then iteratively reducing the number of states until some minimal
number of states $K_\mathrm{min}$ is reached.
Each iteration involves either merging the two ``closest'' states \emph{or}
deleting the ``smallest'' state, and then re-running
HMMGLasso with one fewer state, using estimates from the previous step
as initialization.
This scheme is summarized in
Algorithm~\ref{alg:greedybw}.

\begin{algorithm} % enter the algorithm
% environment
\caption{Greedy Backward Pruning with HMMGLasso}
\label{alg:greedybw}
\begin{algorithmic}[1] % give the algorithm a caption
% and a label for \ref{} commands later in the document
% enter the algorithmic environment
\State\textbf{Input} $K_{\mathrm{min}}$ and $K_{\mathrm{max}}$.
Initialization of
$\Upsilon^{(K_{\mathrm{max}})} = \{(\mathrm{u}_k(t))_{k=1,\ldots,K_\mathrm{max},t
\in
\mathcal{T}},  \Pi, \pi\}.$%\vspace{0.1cm}
\State Fit \textbf{HMMGLasso} and obtain: $\hat\Xi^{(K_\mathrm{{max}},\lambda_\mathrm{uni})}\leftarrow\textbf{HMMGLasso}(K_\mathrm{max},
\lambda_\mathrm{uni}, \Upsilon^{(K_{\mathrm{max}})})$.\vspace{0.1cm}
\State Set $\kappa=K_\mathrm{max}$.
\While{$\kappa> K_{\mathrm{min}} $}
\State\textbf{Merge \emph{Or} Delete}

Compute merged/deleted initial conditions:
$\Upsilon_{\mathbf{mer}}$ and $\Upsilon_{\mathbf{del}}$.

Compute $\Xi_{\mathbf{mer}}\leftarrow\textbf{HMMGLasso}(\kappa
-1,\lambda
_\mathrm{uni},\Upsilon_{\mathbf{mer}})$

Compute
$\Xi_{\mathbf{del}}\leftarrow\textbf{HMMGLasso}(\kappa-1,\lambda
_\mathrm{uni},\Upsilon_{\mathbf{del}})$.
\State\textbf{Update:}

Set $\kappa\leftarrow\kappa-1$.

Set $\hat{\Xi}^{(\kappa,\lambda_\mathrm{uni})}\leftarrow\Xi_{\mathbf{mer}} $
\emph{if}
$\mathcal{C}(\Theta_{\mathbf{mer}})<\mathcal{C}(\Theta_{\mathbf{del}})$.

Set $\hat{\Xi}^{(\kappa,\lambda_\mathrm{uni})}\leftarrow\Xi_{\mathbf{del}} $
\emph{if}
$\mathcal{C}(\Theta_{\mathbf{del}})<\mathcal{C}(\Theta_{\mathbf{mer}})$.

\EndWhile
\State\textbf{Set:} $\hat{K}_{\mathrm{opt}}=\argmin_\kappa
\mathcal{C}(\hat{\Theta}_{\kappa,\lambda_\mathrm{uni}})$.
%% \State Fine-Tuning (optional):
%% Re-run \textbf{HMMGLasso} ($K$ fixed at
%% $K_{\rm{opt}}$) on a $\lambda$-grid $\Lambda=\{\lambda_m
%% Set $\lambda_\mathrm{opt}=\argmin_{\lambda\in
%% \Lambda}\mathcal{C}(\hat{\Theta}_{K_\mathrm{opt},\lambda})$.
\State\textbf{Output} final estimates: $\hat{\Theta}_{K_\mathrm{opt},\lambda_\mathrm{uni}}$.
\end{algorithmic}
\end{algorithm}

We give now a definition of ``smallest'' state and ``closest'' states
and describe the ``merge'' and ``delete'' operations in detail. Let
$\hat\Theta_K$ be the current estimate for~$K$ states. The merge
operation consists of
detecting the two closest states $k_1$ and $k_2$ defined as
\[
(k_1,k_2)=\argmin_{k,k'\in\{1,\ldots,K\}}\D_s(\hat
\theta_k\Vert \hat \theta_{k'}),
\]
where $\D_s(\hat\theta_k\Vert \hat\theta_{k'})$ is the symmetric
Kullback--Leibler divergence given by
\begin{eqnarray*}
&&\D_s(\hat\theta_k\Vert \hat\theta_{k'})\\
&&\qquad=
\operatorname{tr}\bigl\{(\hat{\Sigma}_k-\hat{\Sigma}_{k'}) \bigl(
\hat{\Sigma}^{-1}_{k'}-\hat{\Sigma}^{-1}_{k}\bigr)\bigr
\}+(\hat{\mu}_k-\hat{\mu}_{k'})^T\bigl(\hat{\Sigma}
^{-1}_k-\hat{\Sigma}^{-1}_{k'}\bigr) (
\hat{\mu}_k-\hat{\mu}_{k'}).
\end{eqnarray*}

We merge states $k_1$ and $k_2$ into a new state (denoted by
$k_1\cup k_2 $) by forming new initial conditions
for the next run of HMMGLasso with $K-1$ states. In particular, we
compute merged responsibilities as
% States need to be similar in terms of symmetric Kullback-Leibler
% divergence $\D_s(\cdot\Vert \cdot)$ and, simultaneously, their joint
% effective sample size should not be large. This measure was
% introduced in. The authors demonstrate the usefulness of an
% agglomerative EM for selecting the optimal number of clusters in a
% finite mixture context.
% \begin{eqnarray*}
% &\mathrm{u}^{(0)}_{\mathbf{mer} k_1\cup k_2}(t)=\mathrm{\hat
%u}_{k_1}(t)+\mathrm{\hat
% u}_{k_2}(t) \operatorname{and} \mathrm{u}^{(0)}_{\mathbf{mer},k}(t)=\mathrm{
% u}_{k}(t) (\operatorname{for} k\neq k_1\cup k_2)\\
% &\Pi^{(0)}_{\mathbf{mer} \rm k_1\cup k_2,k'}=\hat\Pi_{k_1k'}+\hat
% (\operatorname{for all} k',k \neq k_1\cup k_2) \\
% &\Pi^{(0)}_{\mathbf{mer} k',k_1\cup k_2}=1/(K-1) (\operatorname{for} k'=1,
% \end{eqnarray*}
%
\begin{eqnarray*}
\mathrm{u}_{\mathbf{mer} k_1\cup k_2}(t)&=&\mathrm{\hat u}_{k_1}(t)+\mathrm{\hat
u}_{k_2}(t),
\\
\mathrm{u}_{\mathbf{mer} k}(t)&=&\mathrm{\hat u}_{k}(t)\qquad (\mbox{for } k
\neq k_1\cup k_2)
\end{eqnarray*}
and get a merged transition matrix using updates
\begin{eqnarray*}
\Pi_{\mathbf{mer} \mathrm{k}_1\cup
k_2,k'}&=&\hat\Pi_{k_1k'}+\hat\Pi_{k_2k'}\qquad \bigl(
\mbox{for } k' \neq k_1\cup k_2\bigr),
\\
\Pi_{\mathbf{mer} \mathrm{ k},k'}&=&\hat\Pi_{k,k'}\qquad \bigl(\mbox{for } k',k
\neq k_1\cup k_2\bigr),
\\
\Pi_{\mathbf{mer} k',k_1\cup k_2}&=&1/(K-1)\qquad \bigl(\mbox{for } k'=1,\ldots,K-1
\bigr).
\end{eqnarray*}

All these operations are based on the relation
$\PP(S_t = k_1\cup S_t = k_2|\cdot) = \PP(S_t = k_1|\cdot)+\PP(S_t =
k_2|\cdot)$.

The delete operation simply discards the smallest state according
to\break  $\min_{k\in\{1,\ldots,K\}} \hat{\pi}_k$.
Initial conditions $\mathrm{u}_{\mathbf{del}},
\Pi_{\mathbf{del}}$ arising from deleting a state are derived by
omitting the corresponding
row/column of $\hat{\mathrm{u}}, \hat{\Pi}$ and renormalizing these
quantities such that rows sum up to one.\vadjust{\goodbreak}

Note that the Greedy Backward Pruning algorithm needs to be initialized
only once, namely, at
$K_\mathrm{max}$. % The idea is, that at $K_\mathrm{max}$ the surface of the
% penalized neg. log-likelihood function is unspecific and close to
% uniform (note that at the
% limit where $K_\mathrm{max}=n$ the likelihood puts weights $1/n$ on each
% observation). As a consequence HMMGLasso is insensitive to
% initialization at $K_\mathrm{max}$ and
%We start with uninformative initialization of
%$\mathrm{u}_k(t)$ by drawing from a multinomial distribution.
Further, we note from Algorithm~\ref{alg:greedybw} that we decide
between the
``merging'' and ``deleting'' operations based on the model selection
criterion, that is, if initial conditions obtained from merging leads to
an estimate with smaller criterion $\mathcal{C}$,
we choose that solution, otherwise we take the solution obtained from the
``delete'' operation.
As demonstrated in the examples below, Greedy Backward Pruning with
only a single initialization at large $K_\mathrm{max}$ yields remarkably
good estimates in the unknown $K$ case. Our procedure originates from
the algorithms proposed in \citet{Figueiredo1999}, \citet{Figueiredo2000}
and \citet{Bicego2003}.
Our empirical results below echo the findings of these authors that
Greedy Backward Pruning-like approaches
can confer robustness to initialization.

\section{Examples}\label{sec:numexp}
%We present in turn results from simulation studies in which all
%data-generating parameters are known and
%s3.1 #&#
\subsection{Simulation studies}\label{s3.1}
In this section we describe data-generating models that we use for
simulation examples.
We consider the following: %the following data-generating models:
\begin{longlist}[\textit{Model} 1.]
\item[\textit{Model} 1.] $K_\mathrm{true}\in\{2,4,6\}, n=2000, p=10$, ($n/p$-ratio=200).

Transition matrix. $\Pi_{kk'}=0.1\gamma$ and $\Pi_{kk}=0.9\gamma$,
where $k,k'\in\{1,\ldots,K_\mathrm{true}\}$ and $\gamma$ is chosen such
that $\sum_{k'=1}^{K_\mathrm{true}}\Pi_{kk'}=1$.

Means $\mu_{k}, k=1,\ldots,K_\mathrm{true}$. Each state has $p/K_\mathrm{true}$ nonzero entries with value $(-1)^{k}\alpha/\sqrt{p/K_\mathrm{true}}$. Nonzeros are at different locations for each state.

Concentration matrix $\Omega_{k}, k=1,\ldots,K_\mathrm{true}$. Each state
has $p$ nonzero (off-diagonal) entries. To reflect the setting in which
states share some aspects of the graphical model structure,
$p/2$ nonzeros are shared between all states, whereas the other $p/2$
nonzeros are at different locations for each state. Concentration
matrices are generated as in \citet{rothman2008spi} but standardized to
have unit diagonal entries.

\item[\textit{Model} 2.] As model 1 but with $p=75$, ($n/p$-ratio${}=26\ 2/3$).

\item[\textit{Model} 3.] As model 1 but with $n=1000$ and $p=100$, ($n/p$-ratio${}=10$).

\item[\textit{Model} 4.] $K_\mathrm{true}\in\{2,4,6\}, n=5000, p=50$.

Transition matrix. $\Pi_{kk'}=0.1\gamma$, $\Pi_{kk}=0.9\gamma$ for
$k\neq K_\mathrm{true}$; $\Pi_{K_\mathrm{true},k'}=1/K_\mathrm{true}$ ($k'\in\{
1,\ldots,K_\mathrm{true}\})$. Again, $\gamma$ is chosen such that rows sum
up to one.

Means. $(\mu_k)_l=\alpha$ for $l\in\{1,2\}$ and $k\in\{1,2\}$. All
other entries equal zero.

Concentration matrix. For $k=1,2$: $\Omega_{k}=\mathrm{I}_p$.
For $k=3,\ldots,K_\mathrm{true}$: $\Omega_{k}$ has two
nonzero entries, at different locations for each state. Concentration
matrices are standardized to have unit diagonal entries.
\end{longlist}

Ideally we seek methodology that can automatically adapt to both low-
and high-dimensional settings.
Accordingly, models~1, 2 and 3 have the same design but differ with
respect to the $n/p$-ratio. We include the small $p$, large $n$ model~1
as a baseline and to investigate the performance of universal\vadjust{\goodbreak}
regularization in the classical low-dimensional setting. Model 4 is a
challenging problem, similar in terms of $n,p$ to the real, genomic
data example below.
%
%Model 4 has different design, i.e., transition matrix, and
%state-specific parameters are generated in a different way than in
%models 1, 2
%and 3.

%pa3.1.subsubsection.1 #&#
\textit{Experiment \textup{I:} Number of states.}
In this experiment the focus is on state recovery. We explore the
ability to estimate the correct number
of states $K$ and recover the state assignments.
We compare the following methods:
\begin{itemize}
\item HMMGLasso initialized by Kmeans
(\emph{Hmmgl});
\item HMMGLasso with \emph{Greedy Backward Pruning}
(\emph{Bwprun});% (without fine-tuning step)
\item Unpenalized maximum likelihood estimation (MLE)
(\emph{Unpen});
\item MLE with
diagonal restricted covariance matrices (\emph{Diagcov});
\item Model-based clustering via Gaussian mixture models [\emph
{Mclust}; \citet{MclustSoftwareManual2006}].
\end{itemize}

Thus, \textit{Hmmgl} and \textit{Bwprun} are the methods we propose.
Both Hmmgl and Bwprun carry out estimation (for given $K$) using the
penalty and universal regularization via $\lambda_\mathrm{uni}$ that we
put forward above; the former embeds our estimator within a standard,
``brute-force'' exploration of $K$, while the latter uses Greedy
Backward Pruning.

In all numerical experiments we stop the algorithms according to the
rule described in Algorithm \ref{alg1} with $\varepsilon=10^{-3}$ and $\pi_\mathrm{min}=5/n$
(for Unpen we use $\pi_\mathrm{min}=p/n$ to ensure nonsingular
covariance estimates). For each method we use each of
BIC and MMDL as model selection criteria. For Hmmgl, Unpen and Diagcov
we compute estimates for $K=1,\ldots,K_\mathrm{true}+2$ and
pick the number of states minimizing BIC or MMDL. As a reference, we
also cluster the data using the \textbf{R}-package \texttt{mclust}
[\citet{MclustSoftwareManual2006}]. We use the function \texttt{Mclust};
this employs Gaussian mixture models and uses BIC to automatically
select between different covariance structures and numbers of clusters
(we allow $K=1,\ldots,K_\mathrm{true}+2$). We initialize \texttt{Mclust} using
model-based hierarchical clustering with equal spherical
covariances (we note that the default initialization of
\texttt{Mclust}, using hierarchical clustering with unconstrained
covariances, performs worse in the examples below). For more details
see \citet{FraleyRaftery2002}. Specifications of all the methods are
summarized in Table~\ref{tab:exp1}.
%
%t1 #&#
\begin{table}%[h!]
\tabcolsep=2pt
\caption{Methods used in simulation Experiment~I [r.s. stands for random starts]}
\label{tab:exp1}
\begin{tabular*}{\textwidth}{@{\extracolsep{\fill}}lccc@{}}
\hline
\textbf{Method} &\textbf{Selection criterion} $\bolds{\mathcal{C}}$&\textbf{Regularization/Constraints}&
\textbf{Initialization}\\
\hline
\textup{Bwprun}&BIC/MMDL&($\operatorname{Pen}_\mathrm{parcor}$, $\lambda_\mathrm{uni}$)& KM (100 r.s.) at $K_\mathrm{max}=15$\\
\textup{Hmmgl}&BIC/MMDL&($\operatorname{Pen}_\mathrm{parcor}$, $\lambda_\mathrm{uni}$)& KM (100 r.s.)\\
\textup{Unpen}&BIC/MMDL&No constraints& KM (100 r.s.) \\
\textup{Diagcov}&BIC/MMDL&Diagonal covariances& KM (100 r.s.) \\
\textup{Mclust}&BIC& Various covariance structures& Hierarchical
clustering\\
&&[{see \citet{FraleyRaftery2002}}]\\
\hline
\end{tabular*}
\end{table}

We generated 50 data sets from each of models 1--4 with
$\alpha=2$ and report for all methods number of
selected states and adjusted Rand index (this quantifies the extent to
which estimated state assignments agree with true state membership).
The results for
models~3 and~4 are
summarized in Figures~\ref{fig:model3} and \ref{fig:model4}; Figures~S2
and~S3 in the supplementary
material [\citet{supp}] show results for models 1 and 2.

%f1 #&#
\begin{figure}

\includegraphics{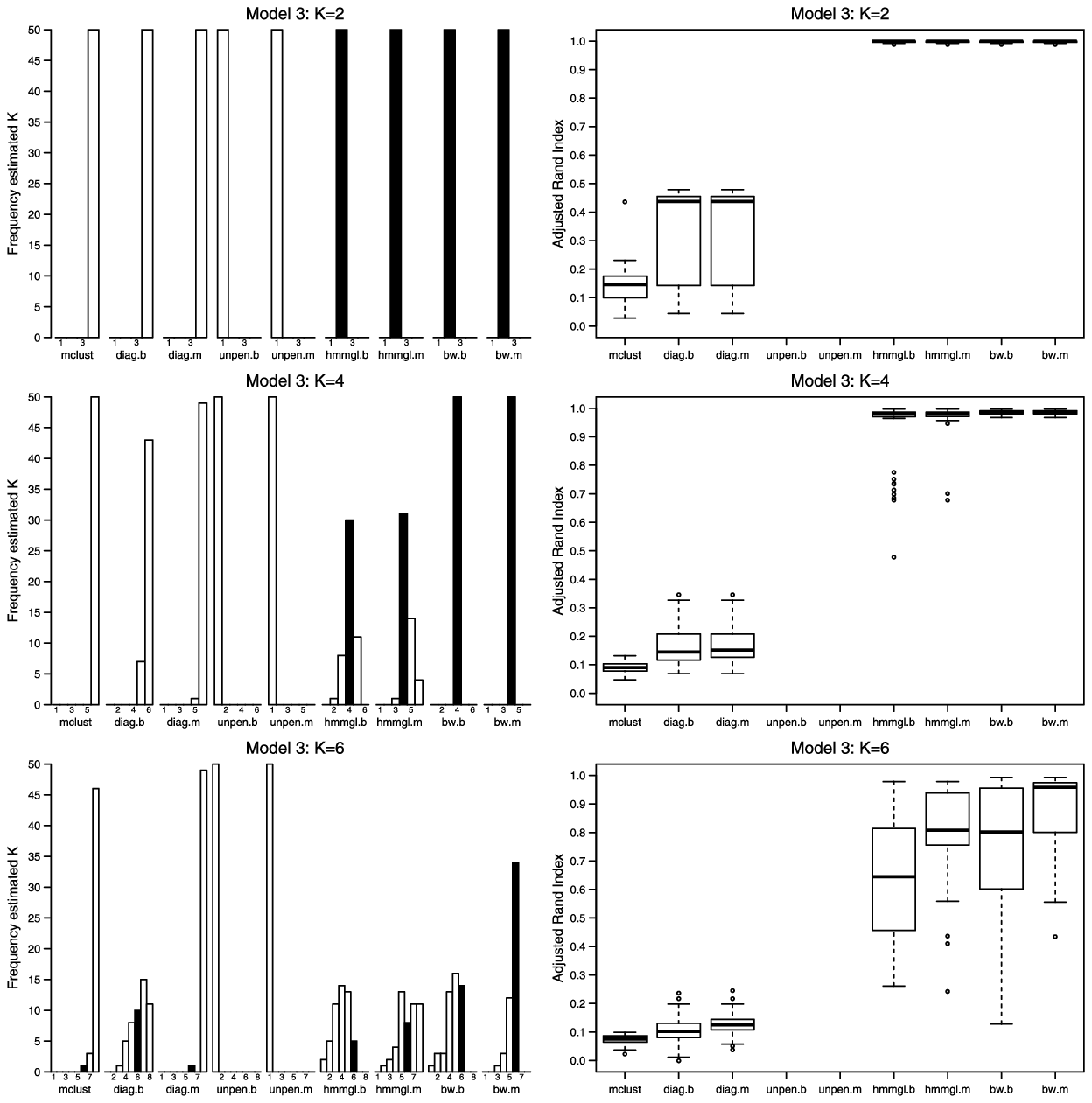}

\caption{Simulation model 3 ($p=100,n=1000$), number of states and
state assignments. Left panels: frequency of estimated number of states;
in each case the correct number of states (i.e., number of states in
data-generating model)
is indicated in black. Right panels: adjusted Rand
index with respect to true state assignments. [Legend: Results for
Mclust (\emph{mclust}), MLE with diagonal covariance
matrices (\emph{diag}), MLE (\emph{unpen}) and Greedy Backward
Pruning (\emph{bw}) are shown.
The extensions ``.b'' and ``.m'' stand for BIC and MMDL, resp.]}
%% eigentliche Bild-Legende.
\label{fig:model3}
\end{figure}

%f2 #&#
\begin{figure}

\includegraphics{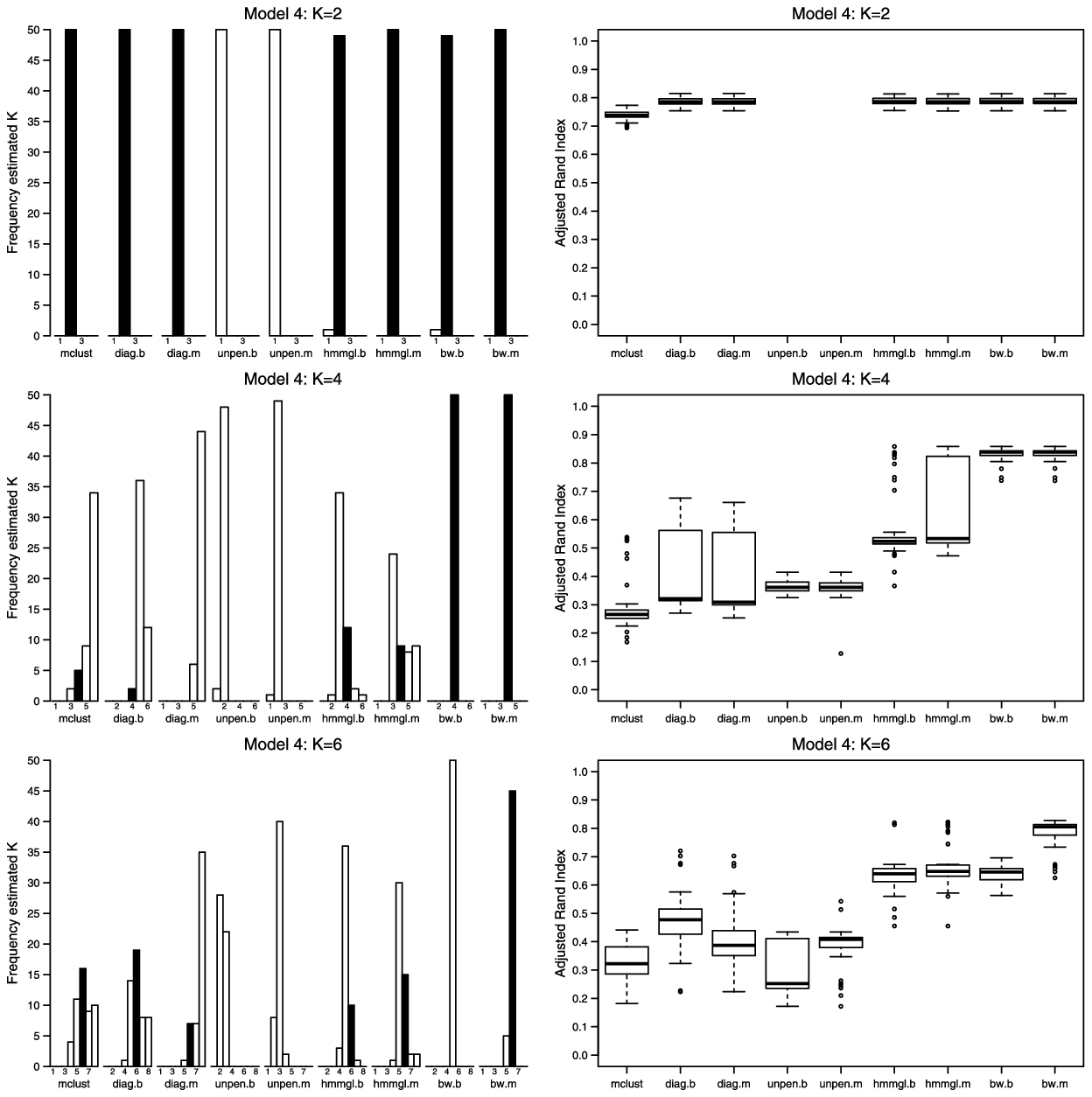}

\caption{Simulation model 4 ($p=50,n=5000$), number of states and
state assignments. Left panels: frequency of estimated
number of states (as in Figure \protect\ref{fig:model3} correct number
of states indicated in black). Right panels: adjusted Rand
index with respect to true state assignments. [Legend: Results for
Mclust (\emph{mclust}), MLE with diagonal covariance
matrices (\emph{diag}), MLE (\emph{unpen}) and Greedy Backward
Pruning (\emph{bw}) are shown. Extensions ``.b'' and ``.m''
stand for BIC and MMDL, resp.]}
% eigentliche Bild-Legende.
\label{fig:model4}
\end{figure}

In nearly all settings Diagcov is unable to recover the correct
number of states and performs poorly in terms of adjusted Rand
index. This is not surprising as Diagcov imposes incorrect model
assumptions. Only in model~3 with $K_\mathrm{true}=2$, where for both states
the data generating covariance matrices are diagonal, does Diagcov perform
well. MLE without penalization
(Unpen) does well only in the low-dimensional model~1. Both the
proposed methods (Hmmgl and Bwprun) greatly outperform the
other methods in models 2--4.
This supports the notion that
regularization can be useful even when sample size $n$ is seemingly large.

%The proposed HMMGLasso imposes in a fully automated way implicit model
%constraints in order to prevent overfitting.
%(due to possible small effective sample size compared to large
%parameter space) while not imposing wrong model constraints.
HMMGLasso
also works well in model~1 with large $n$ and very small $p$, a scenario
where no constraints are necessary. This demonstrates that the adaptive strategy
and universal regularization can be applied without any hand tuning
also in the low-dimensional setting.
We also read off from Figures~\ref{fig:model3}--\ref{fig:model4}
(see especially scenarios with $K=6$) the substantial improvement of Greedy
Backward Pruning relative to HMMGLasso, despite the fact that the
latter carries out essentially a brute-force search over $K$. Also, the
use of
MMDL further improves performance (it never performs worse than BIC).
Especially in tough and very
high-dimensional scenarios (models~3 and~4 with $K=6$), MMDL seems
to perform better.

\begin{figure}

\includegraphics{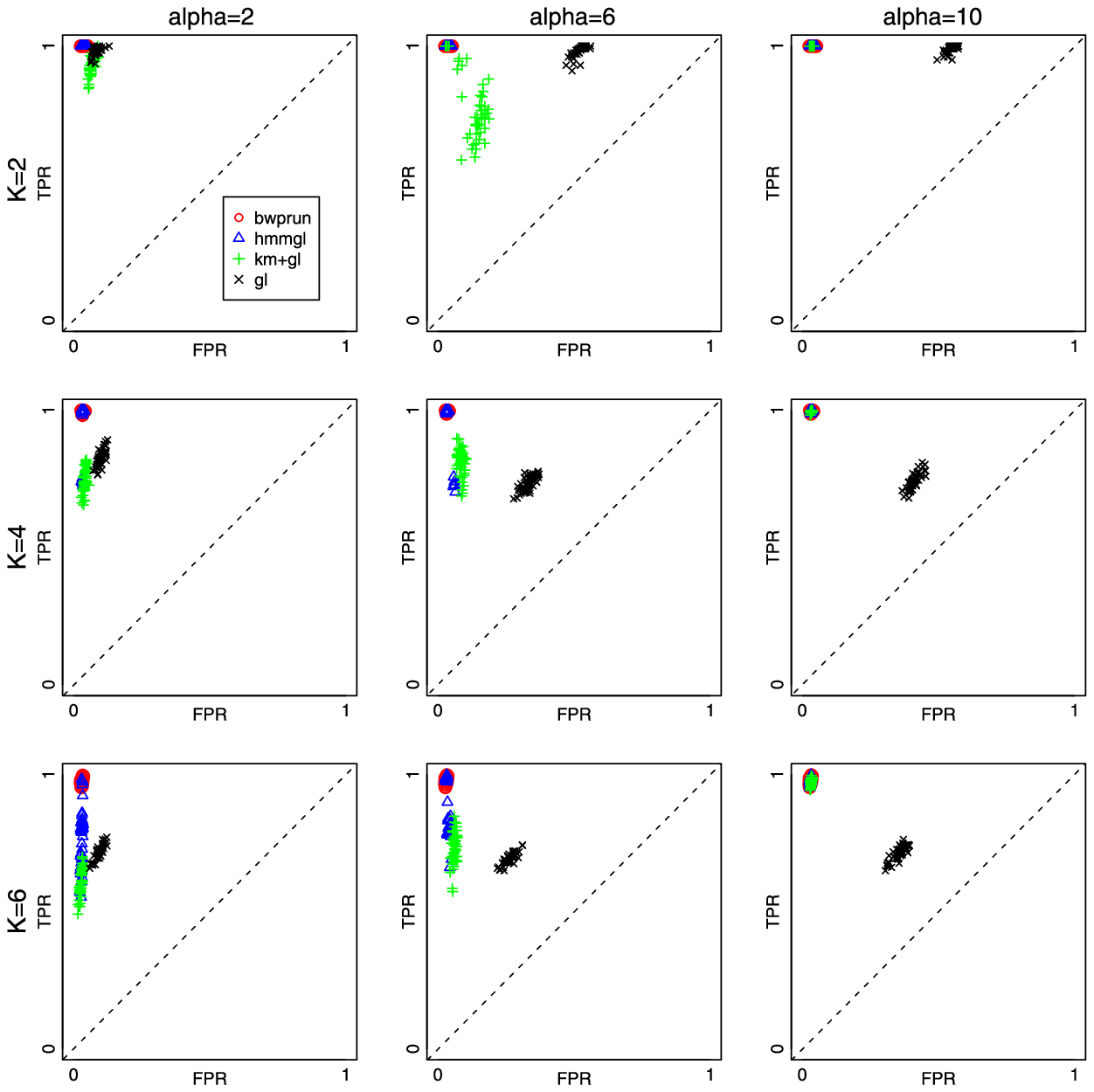}

\caption{Simulation experiment II, graphical model estimation.
Comparing estimated state-specific conditional independence graphs
against the data-generating graphs gave true positive and false
positive rates with respect to edges in the graphs (TPR and FPR,
resp.). We show TPR plotted against FPR with $K\in\{2,4,6\}$,
$\alpha\in
\{2,6,10\}$ for model 3. %(results shown for the most challenging Model
%4, other models appear in SI).
[Legend: Results for Greedy Backward Pruning (\emph{bwprun}),
HMMGLasso (\emph{hmmgl}), Kmeans clustering with cluster-wise
Graphical Lasso
(\emph{km${}+{}$glasso}) and
Graphical Lasso applied to nonclustered data (\emph{glasso}) are shown.]}
%[Legend: Results for Greedy Backward Pruning \emph{without}
% fine-tuning (\textbf{bw}), Greedy Backward Pruning \emph{with}
%fine-tuning
% (\textbf{bw+finetune}), Kmeans clustering with cluster-wise glasso
% (\textbf{km+glasso}) and
% glasso applied to nonclustered data (\textbf{glasso}) are shown.]
% eigentliche Bild-Legende.
\label{fig:exp2roc}
\end{figure}

\textit{Experiment II\textup{:} Graph structure}.
In this experiment we focus on recovering state-specific graphical
model structure. We consider model~3 with $K_\mathrm{true}\in\{2,4,6\}$ and
$\alpha\in\{2,6,10\}$. We compare Greedy Backward Pruning,
HMMGLasso ($K=K_\mathrm{true}, \operatorname{Pen}_\mathrm{parcor}, \lambda_\mathrm{uni}$), Kmeans
(with number of clusters set to $K=K_\mathrm{true}$) followed by
estimating cluster-specific inverse covariance matrices using
Graphical Lasso, and Graphical Lasso
using all samples (no state assignment or clustering). In Figure~\ref{fig:exp2roc} True Positive Rate (TPR; with respect to edges in the
data-generating graph) is plotted
against the corresponding False Positive Rate (FPR) for all
combinations of $K$ and
$\alpha$ and different methods. We note that Greedy Backward Pruning
consistently selects the correct number of states in all scenarios
except in $(K_\mathrm{true},\alpha)=(6,2)$ where it chooses
$K$ correctly in 36 out of 50 data sets.

%f3 #&#

%Greedy Backward Pruning with and without fine-tuning both perform very
%well
%in terms of TPR and FPR. We notice that fine-tuning step leads only to
%a small improvement.
Greedy Backward Pruning performs well in terms of TPR
and FPR. It is noteworthy that universal regularization using
$\lambda_\mathrm{uni}$ gives consistently good results under a range of
conditions. We
see that HMMGLasso exhibits a smaller true positive rate in the most challenging
$K_\mathrm{true}=6$ case.
For $\alpha=2$ Kmeans in combination with
GLasso performs much worse, in particular in terms of TPR. For larger
$\alpha$'s (and therefore with increased
information about state-assignment in the means) TPR and FPR of Kmeans
improves. Finally, GLasso applied to all data without any clustering
leads to very poor performance (this is likely a consequence of
Simpson's paradox).

\subsection{Application to genomic data}\label{sec:application}
We consider genome-wide binding data for 53 proteins in
the \textit{Drosophila} cell line Kc167
[data from \citet{filion2010}].\vadjust{\goodbreak} \citet{filion2010} represents an
important step forward in the genome biology of
\textit{Drosophila}, showing how multivariate data can reveal
protein-DNA binding patterns that depend on genome region.
Here, we use this data set to test our HMM methodology. The data set
offers a number of advantages for our purposes.
First, the
coverage of a relatively large number of proteins ($p=53$) in the full
data gives a high-dimensional example from current genome biology.
Second, the abundance of data ($n=33\mbox{,}632$ for chromosome 2L and
$n=32\mbox{,}791$ for chromsome~2R) allows
fully held-out validation on a large test set (we use the latter half
of chromsome 2R, giving $n_\mathrm{test}=16\mbox{,}396$) as well as exploration
of the effect of (training) sample size.
Finally, although substantive biological questions are beyond the scope
of this paper,
several open questions concerning genome organization in \textit
{Drosophila}, including
the likely number of genome regions, and the possibility of
region-specific protein--protein interplay, help to
motivate the methodological questions we address here.

\citet{filion2010}
identified regions of the genome
by fitting a HMM (using classical, unpenalized estimation) to
reduced-dimension data.
Dimensionality reduction was carried out using principal component
analysis (PCA) as a preprocessing step, with the HMM fitted to the
first three
principal components.
Such approaches are currently widely used in genome biology.
By looking at
principal components, \citet{filion2010} suggested a model with five
states (corresponding to different chromatin types). They further noted that
these five states are marked by enriched binding of the proteins
\emph{HP1, PC, H1, BRM} and \emph{MRG15} and that a 5-state HMM using
only the five marker proteins as an input recapitulates $85.5\%$ of
the original state classification.

We investigated performance in a held-out predictive sense by training
on the
first $n_\mathrm{train}=500, 1000, $ $2000, \ldots, 5000$ observations of
chromosome 2L and then reporting the test log-likelihood obtained from
the second half of chromosome 2R ($n_\mathrm{test}=16\mbox{,}396$).
As above, we compare
HMMGLasso (\emph{Hmmgl}),
Greedy Backward Pruning
(\emph{Bwprun}),
unpenalized MLE (\emph{Unpen}) and MLE with
diagonal covariance matrices (\emph{Diagcov}). Additionally, we include
a five-state MLE using only the five marker proteins reported by \citet
{filion2010}
(\emph{Marker}).
For Hmmgl, Unpen and Diagcov the number of
states is determined by exploring different $K$'s in a forward stepwise
manner. We use MMDL and BIC as model selection criteria. All methods are
initialized by Kmeans with initial centroids obtained using
hierarchical clustering; this renders the overall analysis
deterministic by removing variability due to random initialization of Kmeans.

\begin{figure}

\includegraphics{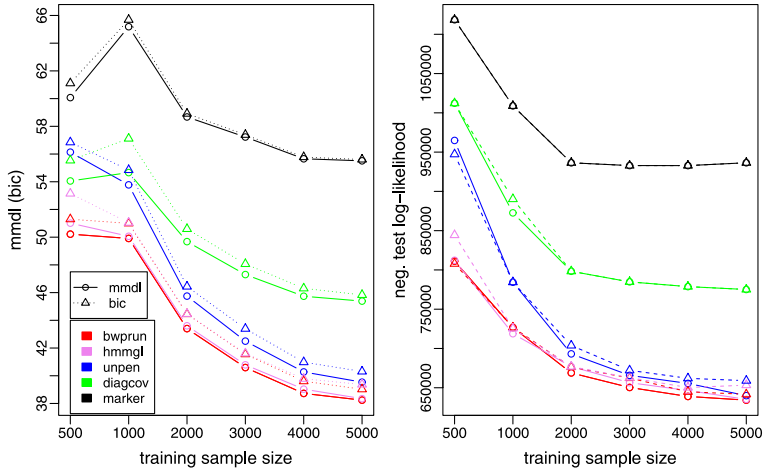}

\caption{Genomic data, MMDL(BIC) and predictive performance. Models
were fitted to protein binding data from Filion et~al. (\citeyear{filion2010}) (see text
for details) and tested on held-out data from the same study.
Left panel: MMDL(BIC)-scores (scaled by $n_\mathrm{train}$) for different
methods trained on the first $n_\mathrm{train}=500, 1000, \ldots, 5000$
observations of chromosome 2L. Right panel: negative
test log-likelihood evaluated on a test set (second half of chromosome
2R; training data is from parts of chromosome 2L). [Legend: Greedy
Backward Pruning (\emph{Bwprun}); HMMGLasso (\emph{Hmmgl});
Unpenalized MLE (\emph{Unpen}); MLE with diagonal restricted
covariance matrices
(\emph{Diagcov}); Five-state MLE using only marker proteins (\emph
{Marker}).]}
\label{fig:testloglik}
\end{figure}

\begin{figure}

\includegraphics{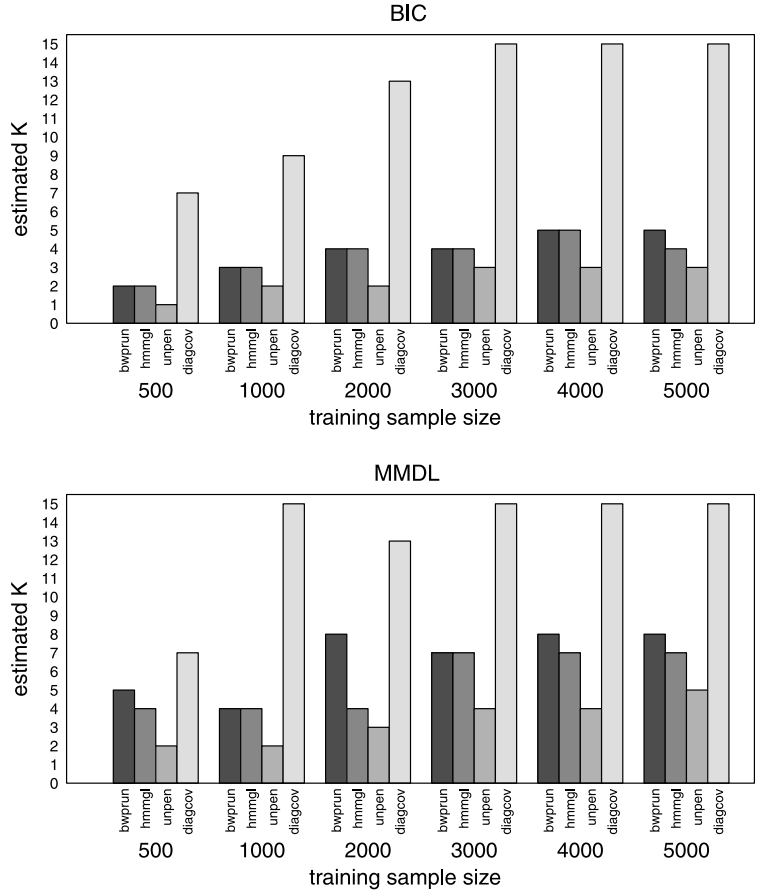}

\caption{Genomic data, number of states. Number of states selected (at
various training sample sizes) by Greedy Backward
Pruning (\emph{Bwprun}), HMMGLasso (\emph{Hmmgl}), unpenalised MLE
(\emph{Unpen}) and MLE with diagonal restricted covariance matrices
(\emph{Diagcov}). All methods are trained on parts of chromosome 2L
and use MMDL or BIC as the model selection criterion.
The number of states in Hmmgl, Unpen and Diagcov are determined by a
forward stepwise selection.}% eigentliche Bild-Legende.
\label{fig:nostates}
\end{figure}

Figure~\ref{fig:testloglik} shows the MMDL(BIC)-scores
(scaled by $n_\mathrm{train}$) and the negative test log-likelihood as a
function of $n_\mathrm{train}$. Figure~\ref{fig:nostates}
depicts the selected number of states for each method and training
sample size. Overall, we notice that MMDL (BIC) and test log-likelihood show
similar patterns for different methods and different sample
sizes. Bwprun and Hmmgl greatly outperform Marker and Diagcov. This
provides a topical example where a multivariate view (using all
variables and modeling also state-specific covariances) improves
out-of-sample predictive performance.
The predictive gain of penalization
compared to unpenalized MLE for moderate $n/p$-ratios is also
noteworthy. As expected, the performance of Unpen in terms of MMDL
(BIC) and
test log-likelihood approaches the penalized methods with increasing
sample size. However, in terms of number of states
(Figure~\ref{fig:nostates}), the estimates are very different
even for large $n_\mathrm{train}$, that is, penalization typically leads to
more states than unpenalized MLE.
This illustrates that the prediction-optimal number of states depends
on the estimation procedure employed: regularization allows estimation
for a greater number of states. If state-specific estimates have
scientific relevance, this property can be important, since, due to
Simpson's paradox, estimates for finer state distinctions (larger~$K$)
cannot, in general, be recovered from coarser models (smaller~$K$).
We return to the question of exploration of number of states in the
\hyperref[sec4]{Discussion} below.

%f4 #&#

%f5 #&#

We note that for each training sample size $n_\mathrm{train}$ the results
shown in Figures~\ref{fig:testloglik}--\ref{fig:nostates} reflect
performance for a single training sample of the specified length. For
completeness, Figure~S4 in the supplementary
material [\citet{supp}] shows performance over 9 different training data
sets of size $n_\mathrm{train}=1000$.

\section{Discussion}\label{sec4}

We considered penalized estimation in multivariate HMMs, including, in
particular,
the case of high dimensions and state-specific graphical models.
%We proposed pebalized
%
%We demonstrated that the approach we propose yields substantive gains
%in estimation compared with both classical, unpenalized estimation,
%and straightforward penalization by incorporation of GLasso into the
%M-step.
As we demonstrated in simulated and real data examples, the methodology
we propose substantially improves upon current practice.
Our results demonstrate the utility of regularization for HMMs, even
when sample sizes are not small.

It is interesting to consider why careful penalization is needed in
HMMs (and related latent variable settings like mixture models).
%% Up to our knowledge, methodology for inference in such models is
%mostly
%% restricted to the very low-dimensional scenarios with only very few
%features $p$
%% compared to a large sample size $n$. This paper goes a step further
%% and includes the
%% high-dimensional view of inference in such models.
In a simple linear model,
as in regression, the ratio
$n/p$ is a measure to distinguish between a low-
and high-dimensional problem. If the ratio
$n/p$ is small, classical least-squares estimation leads to poor
predictive performance due
to a large number of predictors compared to a small sample size.
% In linear regression with $n/p<1$, the maximum likelihood estimator is
% even not properly defined as the design matrix has a low
% rank.
On the other hand, if $n/p$ is large (e.g., $>$20), then, very likely,
least-squares regression performs well.
In HMMs (and mixtures) the situation is more
subtle. It is instructive to consider the ratios
$n_k/p$ ($n_k$
denotes the number of samples belonging to state $k$) as a measure
whether an
inference problem is high-dimensional or not. If for at least one state
this ratio is small, then MLE is likely
to overfit and results in a poor generalization error. A fundamental
problem that we emphasized throughout the paper is the fact that the
ratios $n_k/p$ depend on the number of states $K$ and on the
state-sizes $n_k$, which are themselves usually unknown {a
priori}. So, a seemingly low-dimensional problem with a
large sample size and with a moderate number of features can become a
high-dimensional task in practice, especially if a large number of
states cannot be ruled out {a priori}. In fact, our simulations
illustrate that even when $\min_k n_k/p$ is relatively large, the MLE
can be ill-behaved. For example, in our simulated model~2, with $K=2$,
we have $n=2000$ and $n_k/p > 13$ in each state; nevertheless, the MLE
fails completely to recover correct state assignments [see Figure~S3,
supplementary
material, \citet{supp}].

A straightforward approach to handle inference in high-dimensional
HMMs is to fix constraints on the state-specific covariance matrices
(e.g., assuming diagonal covariance matrices). However, such an
approach leads to poor predictive performance when the assumption is
invalid and precludes discovery of state-specific covariance structure.
As in the genome biology example we considered, such structure may
itself be of scientific interest. We note also that the hidden nature
of the states makes it difficult to test any such model assumption. In
fact, if the covariance matrices of an HMM
with a specific number of states satisfy some constraints, then these
constraints do not necessarily hold for
an HMM with smaller or larger number of states (Simpson's paradox).

Estimation of the number of states in a HMM (or mixture model) remains
challenging.
The backward pruning approach we proposed gives an efficient way to
estimate parameters for a sequence of candidate number of states $K$.
If desired, a single ``optimal'' number of states can then be selected
using model selection criteria, as we demonstrated in the examples above.
Several recent efforts in genome biology have sought to use statistical
criteria to elucidate the number of states in the genome [\citet
{filion2010,ernst2010}] and the methodology we propose can help to
further explore this question in a truly multivariate manner. However,
it is important to emphasize the limitations of model selection
approaches in scientific settings of this kind.
Under model misspecification, in general there is no guarantee that the
correct number of states
will be selected.
To illustrate this effect empirically, we simulated data under model 3,
but with contamination by samples drawn from a multivariate t
distribution [Figure~S5, supplementary
material, \citet{supp}].
We find that although estimation of the number of states holds up well
for lighter tailed contamination, for heavier tails it is demonstrably
inaccurate.
Such behavior is unsurprising, even in the large sample setting, since
under model misspecification we would then expect to recover the model
closest in Kullback--Leibler sense to the data-generating model, which
may not be the model with the scientifically correct number of states.
These observations underline the need for care in scientific
applications where the number of states may have a physical or
biological interpretation and where some degree of model
misspecification is likely unavoidable
[in the related setting of mixture modeling, see, e.g., the
discussion in \citet{hennig}].

In light of the foregoing observations concerning model
misspecification, it is interesting to consider the interplay between
model selection and regularization.
% and the distinction between model selection for scientific
%understanding on the one hand and prediction on the other.
For a given estimator, the optimal number of states is well defined in
a predictive sense as the value that minimizes risk.
From this point of view it is easy to understand why the
prediction-optimal number of states may be higher under regularization
or when more training data are available (see Figure~\ref{fig:nostates}).
For these reasons, when scientific understanding rather than prediction
alone is one of the goals of analysis,
it is not clear whether it is useful to think in terms of a ``correct''
number of states.
Rather, it may be useful to think of estimates $\{ \Theta_K \}$
(obtained, e.g., via backward pruning) as collectively providing
a resource for exploration of a system of interest.

In the context of mixtures, there is a growing literature on penalized
likelihood methods which address the high-dimensional context to some
extent [\citet{khalili,fmrlasso2009,pan,hill2013}]. However, none of
these methods addresses the need to ensure penalties are able to handle
state-specific scaling (that cannot be dealt with by preprocessing) and
size (i.e., unknown at the outset). The selection of the number of
mixture components also remains an open issue in this literature.
%subtle but critical points arising from the hidden nature of mixture
%components: In the classical Lasso or Graphical Lasso it is common
%practice to scale the data-matrix to have unit variance prior to
%estimating parameters. In mixtures (and also HMMs) this is not
%possible, different states will vary in state-size and variables will
%be on a different scale. It is clear that good penalty functions have
%to account for such scaling issues.
%Another important point not addressed so far is the highly sensitivity
%to initialization of EM-type algorithms in fitting penalized HMMs and
%mixtures.
Our approach handles these issues that arise due
to the hidden nature of the states and could be
straightforwardly applied in the mixture model setting. Further
generalization to other latent variable models may also be possible.

In the genome biology example we considered, penalization led to gains
in predictive ability
relative to the MLE and to reduced dimension approaches that have been
used in the literature. This suggests that despite redundancy in
biological signals, a multivariate view can enhance predictive ability.
Further, we were able to learn richer models than are possible using
currently available methods, including estimates of state-specific
graphical model structure.
The latter may shed light on protein--protein interplay that is specific
to genomic region; such interplay has not been investigated to date and
is one focus of our ongoing efforts in this application area.
We used data from \citet{filion2010}; we note that
the main substantive conclusions drawn in that paper are broadly
supported by our analyses and the richer set of states uncovered by our
approach are related to the states they report.
Genomic data sets are becoming increasingly high-dimensional and we
anticipate that the methodology presented here will be useful to
researchers in that field. Beyond biology, potential applications for
high-dimensional HMMs are numerous, including in signal processing and finance.

%For the genome biology example that we considered the statistically
%optimal number of states for our model was XXX; however, further work
%using ancillary, genomic information will be needed to use the
%estimates we obtain for various $K$ to better define chromatin regions
%and genome organization.

%In this example, see Figure~\ref{fig:nostates}, the optimal number of
%states increases with sample size. This fact raises some questions
%about the biological interpretation of the model. The optimal number
%of states is a signature for a model with good predictive performance.
%Such a model in turn can give new insights into the biology of
%chromatin. Obviously, the (statistical) optimal number of states
%cannot be equated with a hypothetically true underlying number of
%(biological) different chromatin states.

We showed that the approaches we put forward for HMMs, including universal
regularization and Greedy Backward Pruning, work well in empirical
examples. However, there remains a need for theoretical investigation
of these ideas.
Our penalty in combination with $\lambda_\mathrm{uni}$ was inspired
by making connections to results obtained for the well-studied Lasso case.
%We explained and motivated the robustness of Greedy Backward Pruning
%to initialization by a rather vague allegory with deterministic
%annealing.
A challenge for future theoretical work is to provide
insight into optimality of these and related approaches
and to establish global convergence properties of
penalized estimation in latent variable settings.

% - general stuff about mixtures being nontrivial critique of existing
%methods.
% - Extensino to mixtures (easy) and general latent variable models\\

%- Theory: our results raise theoretical questions that remain
%unresolved\\

%- applications: better methods open up ability to ask richer
%questions. genome biology - what we're doing

\section*{Acknowledgments} We are grateful to Bas van Steensel and his
lab for introducing us to the genome biology of \textit{Drosophila} and
for a productive, ongoing collaboration and to the Editor and anonymous
referees for their valuable input.

\begin{supplement}[id=suppA]
\stitle{Graphical Lasso with different penalty functions and
supplementary figures}
\slink[doi]{10.1214/13-AOAS662SUPP} %[doi,text={...}] - jei reikia suskaldyti doi
\sdatatype{.pdf}
\sfilename{aoas662\_supp.pdf}
\sdescription{Optimization and
performance of the Graphical Lasso
with the penalty functions $\mathrm{Pen}_{\mathrm
{invcov}}$, $\mathrm{Pen}_{\mathrm {parcor}}$ and $\mathrm{Pen}_{\mathrm
{invcor}}$ introduced in Section~\ref{sec:hmmglasso}. Additional Figures S2--S5 for
Sections~\ref{s3.1}, \ref{sec:application} and~\ref{sec4}.}
\end{supplement}

%
%
% imsref loaded by akundreckaite, 2013-08-23 15:07:12

%

\printaddresses

\end{document}